\documentclass[aps, prl, twocolumn]{revtex4-1}

\usepackage[left=1in, right=1in, top=1in, bottom=1in]{geometry}

\usepackage{mathtools}
\usepackage{amsfonts}
\usepackage[normalem]{ulem}
\usepackage[T1]{fontenc}
\usepackage{dsfont}
\usepackage{array}
\usepackage{color}
\usepackage{braket}
\usepackage{hyperref}

\newcommand{\bibs}{C:/Users/Ethan/Dropbox/References/BibFile}

\begin{document}

\title{Topological control of the nonlinear-optical response of hybrid quantum systems}
\author{Ethan L. Crowell}
\author{Mark G. Kuzyk}
\affiliation{Department of Physics and Astronomy, Washington State University, Pullman, Washington  99164-2814}
\begin{abstract}
We map the topological properties of a one dimensional superlattice to the optical properties of an electronic system.  We find that the nonlinear-optical response is optimized for electrons that live in the transitional morphology between topologically protected edge states and delocalized eigenstates.  This provides a novel means of tuning the nonlinear-optical response of hybrid quantum systems.  We show how these characteristics can be used to mimic saturable absorption and illustrate how `quantum cords' can be used to build an efficient all-optical switch.
\end{abstract}
\maketitle

\section{Introduction}
Optical circuits are characterized by ultra fast operation and low heat dissipation compared to their electronic counterparts.  Optical fibers for transmission applications are ubiquitously used in long haul telecommunications, but their information-carrying capacity is not fully used because of the optical-electrical-optical conversion bottleneck.  Efficient nonlinear-optical materials would make all-optical switches a reality, leading to significant improvements in current optical circuit technology.\cite{werne17.01}

One of the paradigms of all-optical switches which has been the focus of recent work takes advantage of saturable absorption.\cite{kelle96.01,kelle06.01,ono20.01} The underlying mechanism leverages the many electrons that are excited to higher-energy states by the control light, thereby depleting the ground-state population of absorbers and switching the signal light on as the material becomes transparent.  Devices based on refractive index changes, such as the Sagnac switch, are another approach that has been studied as far back as three decades ago.\cite{jinno90.01,avram91.01,gabri91.01,garve96.02}

The successful implementation of this mechanism requires it to be ultrafast and energy efficient. The speed of the switch is constrained by the carrier relaxation dynamics of the material. Namely, it requires that the charge carriers excited by the `Control' beam quickly decay to the ground state when the Control is turned off.

The energy efficiency depends on the nonlinear absorption of the material. If the nonlinear contributions to absorption are small, then the control beam must be more intense to saturate absorption. This requires energies on the order of tens to hundreds of picojoules \cite{li07.01} for switching, which exceeds the energy budget for commercial applications. This has motivated a significant effort to maximize the optical nonlinearities, with particular focus on using local field enhancements to leverage the nonlinear effects.\cite{ren11.02,kaura12.01,minov15.01,lee14.01,valen15.01}

An alternative approach to maximizing the nonlinearities is to enhance the nonlinear susceptbilities themselves. Methods for maximizing molecular (hyper)polarizabilities have been extensively studied.\cite{shafe13.01, kuzyk00.01, kuzyk00.02} Specifically, focus has been placed on finding molecular units whose (hyper)polarizabilities scale most favorably with the spatial extent of the ground state wave-function, and then utilizing molecular synthesis and engineering to scale up to larger materials.\cite{perez16.01,perez16.02,slepk04.01} An ultralarge nonlinear response obtained by such means would mitigate the need for highly intense beams in all-optical switches, as well as enhance the myriad other applicable phenomena in nonlinear optics such as second harmonic generation, third harmonic generation, and the optical Kerr effect.

In this paper we study new degrees of freedom for controlling the nonlinear optical response and investigate the possibility of realizing efficient all-optical switches. To this end, we exploit recent work on nontrivial topological features in one dimensional superlattices.  Section II defines the 1D superlattice and briefly describes the topological properties in certain of the Bloch states.  Section III studies the optical response of a system of fermions in the superlattice while Section IV utilizes topological physics to mimick saturable absorption, and illustrates how this can be used to implement an all-optical switch. Finally, we conclude with a summary of the main results of the paper.

\section{One Dimensional Superlattice and Topological Edge States}
A superlattice in 1+1 dimensions is defined as a one dimensional lattice with a virtual second dimension. Examples of this include the Fermi-Hubbard model with modulated on-site energy \cite{lang12.01}, the Kronig-Penny model with arbitrary scatters \cite{resho19.01}, and non-interacting trigonometric potentials.\cite{zheng14.01} In this paper we realize the superlattice by superposing a Kronig-Penny lattice with a trigonometric potential sublattice. Namely, we consider systems described by the stationary Schr{\"o}dinger equation
\begin{align}
\left(\frac{\hat{p}^2}{2m} + V(x)\right) \psi(x) = E\psi(x),\label{eqn-stationary SE}
\end{align}
with potential energy
\begin{align}\label{eq:PotEng}
V\left(x\right) = V_0 \cos\left(2\pi\frac{x}{x_0} + \phi\right) + h\sum_{n = - \infty}^{\infty} \delta\left(x - nL\right),
\end{align}
where $x_0$ is the period of the sublattice and $L>x_0$ is the period of the Kronig-Penny lattice. Note that if $x_0$ divides $L$, then $L$ is also the period of the total lattice. The periodicity of the potential energy implies that the states will be of the Bloch form, $\psi(x) = e^{iq x}u_q(x)$, where $q$ is the Bloch vector and the functions $u_q(x)$ have the same periodicity as the potential.

The Hamiltonian is periodic in both the Bloch vector $q$ and the lattice phase $\phi$.  As a result, the Brilluoin zone is effectively 2D and forms a torus.  The lattice phase is then a Bloch vector in a second dimension.

We solve Eqn.~\ref{eqn-stationary SE} for values of crystal momenta $q$ and lattice phase $\phi$ spanning the first Brilluoin zone. The results are single-particle Bloch state vectors $\Ket{n  (q,\phi)}$ indexed by parameters $q$ and $\phi$ with energies $E_{n}(q,\phi)$. Appendix A details the numerical methods used to solve Equation~\ref{eqn-stationary SE} with the potential energy given by Equation~\ref{eq:PotEng}.

\begin{figure}
\centering
\includegraphics[width=\columnwidth]{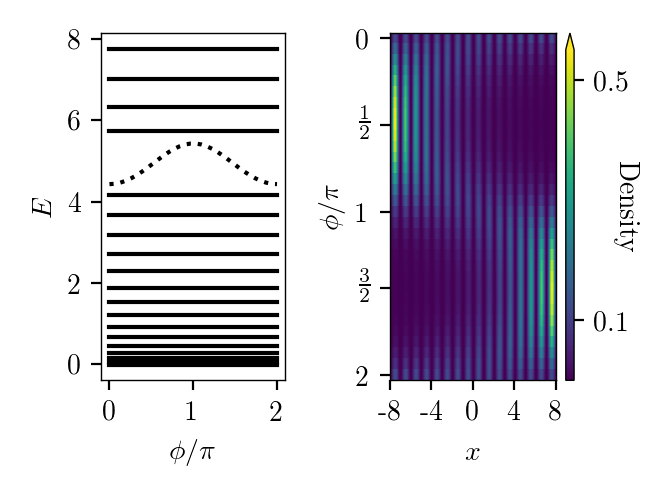}
\caption{(a) The energy spectrum as a function of phase shift for $T=1\, a_0$, $V_0=1.5\,E_H$, and $q=0$. As the phase shift varies smoothly away from $\phi = 0$, the state labeled with a dotted line crosses through the forbidden region and joins the higher band at $\phi=\pi$. (b) The probability density for state $\Ket{16(0,\phi)}$. The state energy lives in the band gap and the wavefunction is well localized near an edge of the box.}
\label{fig-edge state}
\end{figure}

The energy spectrum for $V_0 = 1.5\, E_H$, $x_0=1\, a_0$, and $L=16\,a_0$ ($a_0$ is the Bohr radius and $E_H$ the hydrogen atom binding energy) is shown in Fig.~\ref{fig-edge state}(a) for fixed Bloch vector $q = 0$. We note that we have taken the strength of the Kronig-Penny scatterers, $h$ to be infinite, so each Kronig-Penny cell is essentially a particle in a box.  We first notice that the state $\Ket{16(0,\phi)}$ (denoted with a dashed line) is clearly seen to occupy the band gap as the phase of the lattice is varied between $0$ and $2\pi$. In Fig.~\ref{fig-edge state}(b) we plot the density $|\Braket{x|16(0,\phi)}|^2$ within a single Kronig-Penny cell. We see that the density becomes localized at the boundary of the cell when the state is in the middle of the band gap. Furthermore, the velocity of the edge states is given by $\partial E/\partial \phi$, so from Fig~\ref{fig-edge state}(a) we find that the edge states corresponding to $\phi=\pi/2$ with density localization on the left side of the box and $\phi=3\pi/2$ with localization on the right have opposite velocities.  This gives a certain handedness -- or chirality -- to the band, a feature noted by Reshodko \textit{et al} in the arbitrary Kronig-Penny model.\cite{resho19.01}

The appearance of edge states is attributable to nontrivial topological properties. This can be directly verified from the Cern number of the Bloch state, which is a topological invariant that characterizes the system. Specifically, nonzero values indicate a nontrivial topology. The Cern number is the integral of the Berry curvature over the first Brilluoin zone, given by
\begin{align}
c &= \frac{1}{2\pi}\int_\text{BZ}\text{d}q\text{d}\phi \left(\partial_q A_\phi - \partial_\phi A_q\right),\label{eqn-Cern number}
\end{align}
where $A_\mu = i\Braket{\psi(q,\phi)|\partial_\mu|\psi(q,\phi)}$, with $\mu=q,\phi$, is the Berry connection of the Bloch state. Careful attention must be paid to the fact that the problem is defined on a discretized Brilluoin zone. The link variable approach developed by Fukui \textit{et al} \cite{fukui05.01} was developed explicitly for discretized systems and has been successfully implemented by Reshodko and Lang \cite{lang12.01, resho19.01} for similar systems. The Cern number for the set of states given by $\Ket{16(0,\phi)}$ and shown by the curved band in Fig~\ref{fig-edge state} is exactly one, confirming that the observed features are topological in nature.

\section{Nonlinear Optics of the Edge State}
It is interesting to search for effects of the aforementioned topology on other observables of the system. Here, we investigate the nonlinear dynamics of non-interacting electrons in the superlattice with an incident optical field. The number of electrons $N_e$ is chosen such that the Fermi energy is $E_F= E_{15}(0,\phi)$. The first excited state is thus the excitation of an electron to the edge state band. The topological properties of the band make it possible to tune the HOMO-LUMO gap, as well as the transition moment between the ground and first excited states.

The interaction of a system with an electric field is well described by the (hyper)polarizabilities. These arise by expanding the Fourier components of the electric dipole moment in powers of the incident electric field according to
\begin{align}
p^\omega &= \alpha(-\omega;\omega)\mathcal{E}^\omega + \beta(-\omega;\omega_1,\omega_2)\mathcal{E}^{\omega_1}\mathcal{E}^{\omega_2}\nonumber\\
&\hspace{25pt} + \gamma(-\omega;\omega_1,\omega_2,\omega_3)\mathcal{E}^{\omega_1}\mathcal{E}^{\omega_2}\mathcal{E}^{\omega_3}.
\end{align}
The coefficients $\alpha, \beta$ and $\gamma$ are the polarizability, first hyperpolarizability, and second hyperpolarizability, respectively. For energy conservation, we require $\omega=\omega_1+\omega_2$ for $\beta$ and $\omega = \omega_1+\omega_2+\omega_3$ for $\gamma$.

The diagonal tensor elements of the polarizability are given by the SOS expression \cite{boyd09.01, orr71.01}
\begin{align}
\alpha(-\omega;\omega) &= \frac{N_ee^2}{\hbar}\sideset{}{'}\sum_n\left(\frac{x_{on}x_{no}}{\Omega_{no}-i\gamma_{no} - \omega}\right.\nonumber\\
&\left.\hspace{30pt}+ \frac{x_{on}x_{no}}{\Omega_{no} + i\gamma_{no} + \omega}\right),\label{eqn-SOS alpha}
\end{align}
where $\Ket{o}$ is the many-particle ground state in the limit of zero field strength. We have defined the Bohr frequencies $\Omega_{no} = (E_n-E_o)/\hbar$, the transition moments $x_{nm} = \Braket{n|x|m}$, the phenomenological damping factors $\gamma_{no}$, and the frequency of the incident light $\omega$. The prime on the sum indicates that the state $\Ket{o}$ is excluded to avoid secular divergences for static perturbations.\cite{orr71.01, case66.01}

The first hyperpolarizability is similarly given by
\begin{align}
&\beta(-\omega;\omega_1,\omega_2) = \frac{N_ee^3}{\hbar^2}K(-\omega_\sigma;\omega_1,\omega)2)\mathcal{I}_{1,2}\sideset{}{'}\sum_{n,m}\nonumber\\
& \left(
\frac{x_{on}\bar{x}_{nm}x_{mo}}{\left(\Omega_{no}-i\gamma_{no} -\omega_\sigma\right)\left(\Omega_{mo}-i\gamma_{mo}-\omega_1\right)}\right.\nonumber\\
&\left. + \frac{x_{on}\bar{x}_{nm}x_{mo}}{\left(\Omega_{no}+i\gamma_{no}+\omega_\sigma \right)\left(\Omega_{mo}+i\gamma_{mo}+\omega_1\right)}\right.\nonumber\\
&\left. + \frac{x_{on}\bar{x}_{nm}x_{mo}}{\left(\Omega_{no}+i\gamma_{no}+\omega_2 \right)\left(\Omega_{mo}-i\gamma_{mo}-\omega_1\right)}
\right)\label{eqn-SOS beta}
\end{align}
where $\bar{x}_{nm} = x_{nm} - x_{oo}\delta_{n,m}$ and $\mathcal{I}_{1,2}$ directs us to average over all permutations of $\omega_1$ and $\omega_2$. The numerical coefficients $K(-\omega_\sigma;\omega_1,\omega_2)$ are tabulated in the literature\cite{orr71.01}. There is an analogous expression for the second hyperpolarizability $\gamma$. However, due to its substantive complexity, we relegate the expression to Appendix B (see Eqn.~\ref{eqn-SOS gamma}).

\begin{figure}
\centering
\includegraphics[width=\columnwidth]{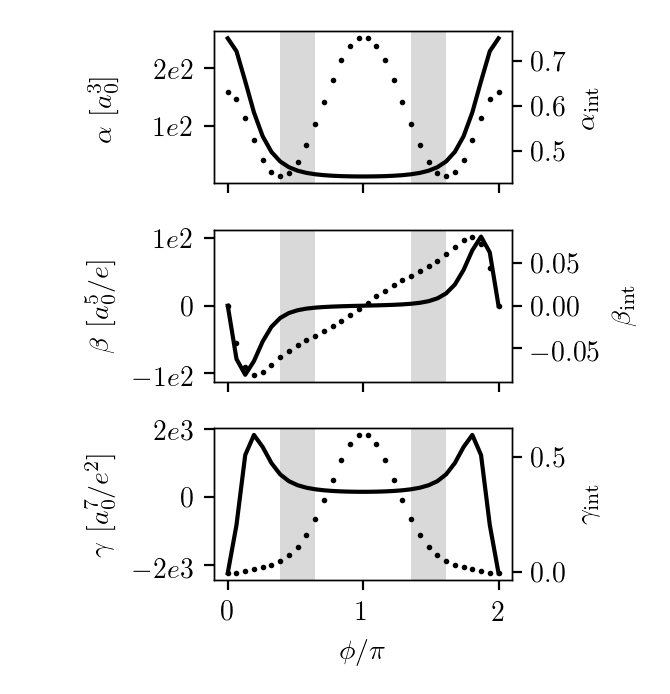}
\caption{(hyper)polarizabilities (solid lines) as function of lattice phase $\phi$. The intrinsic (hyper)polarizabilities (dotted lines) are also shown. The shaded regions show the approximate volume of phase space where the density is localized near the delta-function boundary.}
\label{fig-hyperpolarizabilities}
\end{figure}

The results for static applied fields are shown in Fig.~\ref{fig-hyperpolarizabilities}. The (hyper)polarizabilities are drawn as solid lines and the shaded regions indicate the approximate volume of phase space containing the edge state.  Variation of the phase $\phi$ has a drastic effect on the optical response of the fermionic system. In the localized regime, $\beta$ and $\gamma$ approach zero. This is likely due to the known robustness of edge states to defects and perturbations.

A key figure of merit for the (non)linear optical response of a molecule is the intrinsic (hyper)polarizability. This is defined as the (hyper)polarizability scaled by the fundamental limit imposed by quantum mechanics.\cite{kuzyk13.01, kuzyk00.01, kuzyk00.02} The intrinsic quantities are scale invariant, bounded above/below by 1, and characterize how favorably a given system scales with the spatial extent of the ground state wavefunction.\cite{kuzyk16.01} The intrinsic quantities are shown as the dotted lines in Fig.~\ref{fig-hyperpolarizabilities} . The topological properties of the edge state allow us to vary the intrinsic polarizability by as much as 50\%. What is even more striking is that we are able to continuously tune the intrinsic second hyperpolarizability from 0 to 99\% of the fundamental limit.\cite{lytel17.01}

\begin{figure}
\centering
\includegraphics[width=\columnwidth]{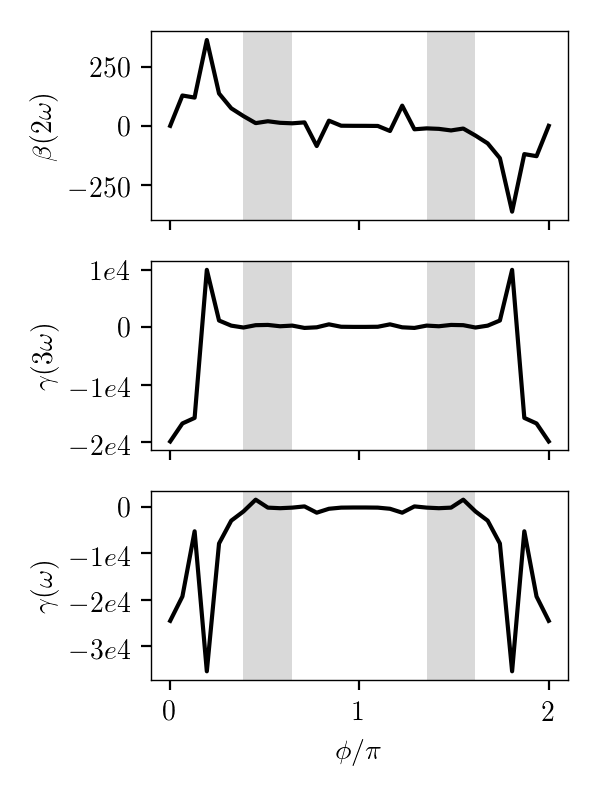}
\caption{Hyperpolarizabilities governing (top) second harmonic generation, (middle) third harmonic generation, and (bottom) the optical Kerr effect for frequency $\omega = 0.7\omega_\text{res}$. All three hyperpolarizabilities are extrema at the same values of lattice shift ($\approx \pi/4$), and nearly zero in the localized regime. }
\label{fig-SHG and THG}
\end{figure}

The dynamic hyperpolarizabilities $\beta\left(-2\omega;\omega,\omega\right)$, $\gamma\left(-3\omega;\omega,\omega,\omega\right)$, and $\gamma\left(-\omega;\omega,\omega,-\omega\right)$ govern second harmonic generation, third harmonic generation, and the optical Kerr effect, respectively. The values of the these coefficients are plotted versus the phase shift in Fig.~\ref{fig-SHG and THG}. The frequency of the incident field was arbitrarily taken to be $70\%$ the resonant frequency at zero lattice shift. We see that the coefficients governing the nonlinear dynamics can be tuned within a range spanning many orders of magnitude. It is also interesting to note that local extrema for each process each occur at the same values of lattice shift $\phi$. We see that the topology of the edge state can also be used to tune the dynamics of the system.

It is interesting to note the similarities between this work and the phase disruption paradigm for increasing the nonlinear-optical response.\cite{lytel15.02} Indeed, we can think of the phase disruption paradigm as a special case of a hybrid edge state; in the former, the position of a single delta in an infinite well is the controllable parameter while in the latter the lattice phase of a periodic structure is controlled.

\section{Simulated Saturable Absorption}
We consider an electric field with frequency $\omega$ incident on an ensemble of 1D quantum systems. Each system is made up of a number of fermions in the superlattice with Fermi Energy $E_F = E_{15}(0,\phi)$. If the field is sufficiently near resonance, then we model each system as having only the two states $\Ket{a}=\Ket{15(0,\phi)}$ and $\Ket{b}=\Ket{16(0,\phi)}$.

We assume that the two states decohere in the characteristic time $T_2$, and that the excited state lifetime is $T_1$. The steady state solution for the coherences in the density matrix can then be used to determine the susceptibility. Defining the Rabi frequency $\Omega = \frac{2|\mu_{ab}|E}{\hbar}$ and the detuning factor $\Delta = \omega - \omega_{ba}$, the real and imaginary parts of the susceptibility can be written as
\begin{align}
\chi' &= -\frac{\alpha_0(0)}{\omega_{ba}/c}\frac{1}{\sqrt{1 + \Omega^2T_1T_2}}\frac{\Delta T_2/\sqrt{1 + \Omega^2T_1T_2}}{1 + \Delta^2T_2^2/(1 + \Omega^2T_1T_2)}\label{eqn-chi real}
\end{align}
and
\begin{align}
\chi'' &= \frac{\alpha_0(0)}{\omega_{ba}/c}\frac{1}{1 + \Omega^2T_1T_2}\frac{1}{1 + \Delta^2T_2^2/(1 + \Omega^2T_1T_2)},\label{eqn-chi imag}
\end{align}
respectively, where
\begin{align}
\alpha_0(0) &= -\frac{\omega_{ba}}{c}\left[N|\mu_{ba}|^2\frac{T_2}{\hbar}\right]
\end{align}
is the unsaturated, line-center absorption coefficient. We emphasize that the above expression holds for arbitrarily large intensities, with the only assumption being that significant coupling only occurs between the two states. For more details on the derivation of Eqns.~\ref{eqn-chi real} and \ref{eqn-chi imag}, see Boyd \cite{boyd09.01}.

\begin{figure}
\centering
\includegraphics[width=\columnwidth]{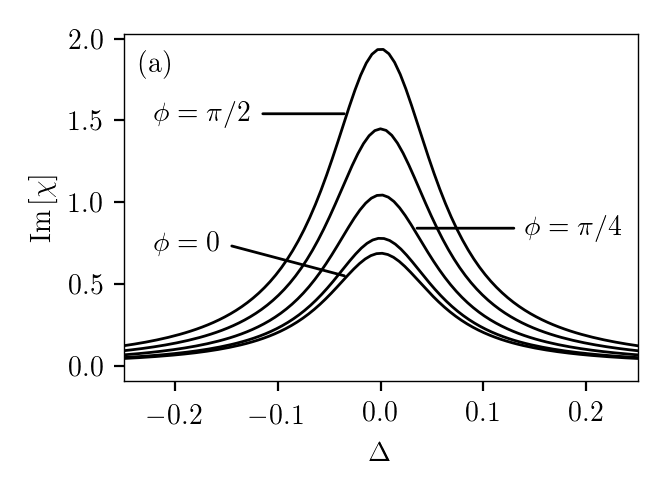}
\includegraphics[width=\columnwidth]{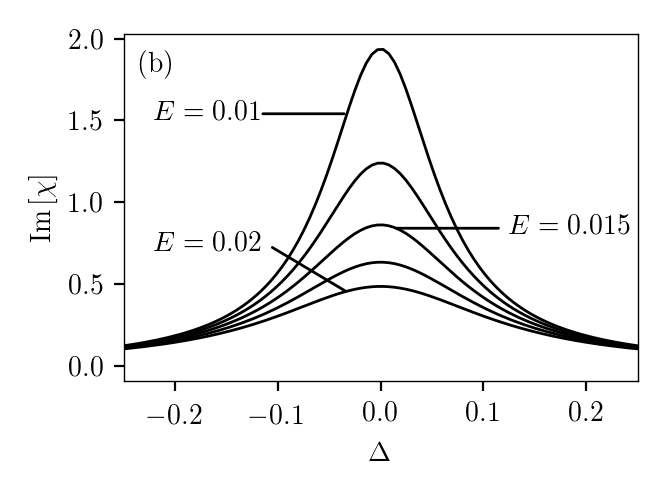}
\caption{(top) Dispersion of the absorption coefficient for select values of lattice phase. We have arbitrarily taken $E=0.01\,E_H/ea_0$ and $T_1=T_2=1000/\omega_{ba}$. (bottom) Absorption coefficient for select values of incident field. The numbers in the chart area are expressed in atomic units. We see that tuning the lattice phase can be used to achieve a decrease in absorption comparable to that of saturable absorption.}
\label{fig-faux saturable absorption}
\end{figure}

We plot the absorption coefficient for select values of phase shift in Fig.~\ref{fig-faux saturable absorption}(a). A phase shift from $0$ to $\pi/2$ results in an approximately $65\%$ decrease in absorption, giving significant control over absorption simply by tuning the phase shift of the lattice potential.  For comparison, we also illustrate standard saturable absorption by plotting the absorption coefficient as the intensity of the field is quadrupled. The decrease in absorption is comparable, suggesting that it is possible to mimick saturable absorption without exciting charge carriers.

One can imagine implementing an all-optical switch within a 3D cylindrical grid of quantum dots using this novel control over absorption. If hopping between adjacent dots is only allowed along the principle axis, then the system constitutes an ensemble of parallel, 1D arrays of quantum dots. We will refer to each 1D array as a `strand' and the ensemble of strands as a `cord'. The superlattice is realized within each strand by an optical standing wave which modulates the on-site energies of the dots. The wavelength of the standing wave is assumed to be sufficiently small so that there are multiple periods within each strand. We assume the beam width of the standing wave is larger than the diameter of the cord so that the lattice is uniform across strands.

\begin{figure}
\centering
\includegraphics[width=\columnwidth]{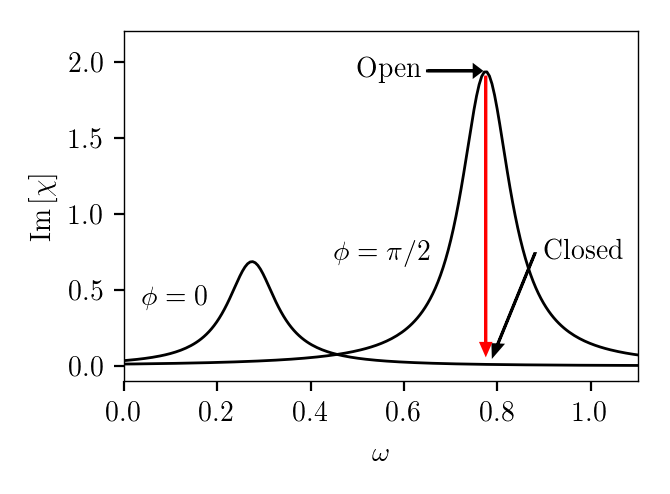}
\caption{Mechanism proposed for an all-optical switch. By switching the lattice phase, the absorption can be decreased by two orders of magnitude.}
\label{fig-switching mechanism}
\end{figure}

The input signal would be an optical field polarized parallel to the cord and nearly resonant with the $\Ket{a}\rightarrow\Ket{b}$ transition at $\pi/2$ lattice shift. The `Off' state corresponds to a $\pi/2$ lattice phase. In this regime the absorption coefficient of the cord is large, so the outgoing signal is arbitrarily small.  To close the switch, we tune the phase of the lattice potential to minimize absorption.  As shown in Fig.~\ref{fig-switching mechanism}, this results in a drastic reduction of the absorption coefficient, thereby allowing an appreciable outgoing signal. Because there is no need to saturate absorption, there is no constraint due to carrier relaxation dynamics.

\begin{figure}
\centering
\includegraphics{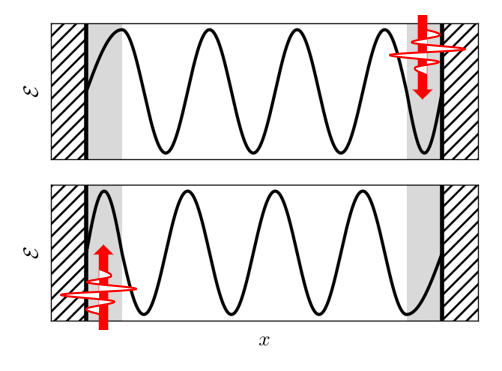}
\caption{An example of how the phase of the lattice can be dynamically tuned using the optical Kerr effect. The walls of the optical cavity (thick vertial lines) are coated with a nonlinear material (gray). Illuminating a film with a pump beam results in a change in the refractive index. By illuminating one film or the other, the phase of the standing wave can be tuned while keeping the wavelength fixed.}
\label{fig-lattice shift mechanism}
\end{figure}

Controlling the phase of the lattice by optical means may require nonlinear effects.  This is not unlike saturable absorption, which is a higher-order nonlinear-optical process.  We discuss one possibility here, though we emphasize there may be many ways of optically controlling the phase.  First, we imagine that the optical standing wave is formed using an optical cavity which is illustrated in Fig.~\ref{fig-lattice shift mechanism}. We assume that both ends of the cavity are coated with a nonlinear material with a large second hyperpolarizability. The optical Kerr effect can be used to tune the refractive index of the coatings using two pump beams. The switch is turned on and off by illuminating one or the other films. Note that in Fig.~\ref{fig-lattice shift mechanism} it is assumed the refractive index of the films can be doubled. However, in practice the films will be many wavelengths thick, and so a small change in the refractive index can result in the needed phase change.

\section{Conclusions}
In summary, we have studied the nonlinear optical properties of an electronic system in a one dimensional superlattice and find that the topology of certain bands provides significant control over the nonlinear optical response of the material. A particularly relevant example of this is the ability to efficiently mimick saturable absorption.  This control can in principle be leveraged in quantum cords to build an all-optical switch that does not require fast carrier relaxation dynamics, but only requires dynamic optical control of the phase.  Similarly, the ultra-fast hyperpolarizabilities can be controlled to optimize applications that require light or voltage-induced phase shifts of light as well light generation through parametric mixing.

The same ideas can be applied to three-dimensional structures that have controllable nonlinear susceptibility tensors that selectively act on the light's polarization.  For example, such control could be used to make polarization independent electro-optic switches.\cite{kuzyk89.03}  This flexibility could give materials and nanostructure designers an additional degree of freedom to control a material's properties.

\section{Appendix A: Numerical Details}
We solve Eqn.~\ref{eqn-stationary SE} using spectral (i.e. Fourier) methods. The benefit of this numerical approach is twofold. First, it naturally accounts for the periodicity of the problem. Second, it drastically reduces errors in the kinetic energy due to discretization of the spatial domain.

To outline the method, we first write the Hamiltonian in the momentum representation,
\begin{align}
H = \frac{\hbar^2k^2}{2m} + V(x).
\end{align}
We then perform the gauge transformation $\psi(x)\rightarrow e^{-iqx}\psi(x) = u_q(x)$. This effectively gives a boost of $\hbar q$ to the mechanical momentum. Under this gauge transformation the Hamiltonian becomes:
\begin{align}
H(q) = \frac{\hbar^2\left(k + q\right)^2}{2m} + V(x).
\end{align}
The eigenstates of this Hamiltonian are of course the periodic functions $u_q(x)$, where the Bloch vector $q$ is treated as a parameter.

Our approach takes advantage of the kinetic energy being diagonal in momentum space. We construct the diagonal matrix
\begin{align}
T_k(q) = \text{diag}\left[-k_\text{max},\dots, k_\text{max}\right]
\end{align}
where $k_\text{max}$ is the Nyquist frequency defined by our spatial grid. The Fourier transform matrix $e^{ikx}$ defines a similarity transformation with which we transform to the position representation,
\begin{align}
T_x(q) = e^{ikx}T_k(q)e^{-ikx}.
\end{align}

We construct the Hamiltonian by broadcasting the potential energy--which is defined on the spatial grid--as a diagonal matrix, and adding it to the kinetic energy matrix $T_x(q)$,
\begin{align}
H(q) = T_x(q) + \text{diag}\left[V(x_1), V(x_2),\dots,V(x_N)\right].
\end{align}
This matrix can then be solved for any given Bloch vector $q$ using standard linear algebra packages (i.e. numpy.linalg in Python).

\section{Appendix B}
The second hyperpolarizability is given by
\begin{widetext}
\begin{align}
\gamma_{ijkr}(-\omega_\sigma;\omega_1,\omega_2,\omega_3) &= -K(-\omega_\sigma;\omega_1,\omega_2\omega_3)\frac{e^4}{\hbar^3}\mathcal{I}_{1,2,3}\times\nonumber\\
&\hspace{-15pt}\sideset{}{'}\sum_{n,m,l}\left\lbrace
\frac{x_{on}\bar{x}_{nm}\bar{x}_{ml}x_{lo}}{\left(\Omega_{no}-i\gamma_{no}-\omega_\sigma\right)\left(\Omega_{mo}-i\gamma_{mo}-\omega_1-\omega_2\right)\left(\Omega_{lo}-i\gamma_{lo}-\omega_1\right)}\right.\nonumber\\
    &+\left.
    \frac{x_{on}\bar{x}_{nm}\bar{x}_{ml}x_{lo}}{\left(\Omega_{no}+i\gamma_{no}+\omega_1\right)\left(\Omega_{mo}-i\gamma_{mo}-\omega_3-\omega_2\right)\left(\Omega_{lo}-i\gamma_{lo}-\omega_3\right)}\right.\nonumber\\
    &+\left.
    \frac{x_{ol}\bar{x}_{lm}\bar{x}_{mn}x_{no}}{\left(\Omega_{n0}-i\gamma_{no}-\omega_1\right)\left(\Omega_{mo}+i\gamma_{mo}+\omega_3+\omega_2\right)\left(\Omega_{lo}+i\gamma_{lo}+\omega_3\right)}\right.\nonumber\\
    &+\left.
    \frac{x_{ol}\bar{x}_{lm}\bar{x}_{mn}x_{no}}{\left(\Omega_{no}+i\gamma_{no}+\omega_\sigma\right)\left(\Omega_{mo}+i\gamma_{mo}+\omega_1+\omega_2\right)\left(\Omega_{lo}+i\gamma_{lo}+\omega_2\right)}\right\rbrace\nonumber\\
    &\hspace{-8em}+K(-\omega_\sigma;\omega_1,\omega_2,\omega_3)\frac{e^4}{\hbar^3}\mathcal{I}_{1,2,3}\sideset{}{'}\sum_{n,m}\left\lbrace
    \frac{|x_{on}|^2|x_{mo}|^2}{\left(\Omega_{no}-i\gamma_{no}-\omega_\sigma\right)\left(\Omega_{mo}-i\gamma_{mo}-\omega_1\right)\left(\Omega_{no}-i\gamma_{mo}-\omega_3\right)}\right.\nonumber\\
    &\left.
    +\frac{|x_{on}|^2|x_{mo}|^2}{\left(\Omega_{no}-i\gamma_{no}-\omega_3\right)\left(\Omega_{mo}+i\gamma_{mo}+\omega_2\right)\left(\Omega_{mo}-i\gamma_{mo}-\omega_1\right)}\right.\nonumber\\
    &\left.
    +\frac{|x_{on}|^2|x_{mo}|^2}{\left(\Omega_{no}+i\gamma_{no}+\omega_\sigma\right)\left(\Omega_{mo}+i\gamma_{mo}+\omega_1\right)\left(\Omega_{no}+i\gamma_{no}+\omega_3\right)}\right.\nonumber\\
    &\left.
    +\frac{|x_{on}|^2|x_{mo}|^2}{\left(\Omega_{no}+i\gamma_{no}+\omega_3\right)\left(\Omega_{mo}+i\gamma_{mo}+\omega_1\right)\left(\Omega_{mo}-i\gamma_{mo}-\omega_2\right)}
\right\rbrace\label{eqn-SOS gamma}
\end{align}
\end{widetext}
The definition of $\bar{x}_{nm} = x_{nm} - x_{oo}\delta_{n,m}$ is the same as for Eqn.~\ref{eqn-SOS beta}, while $\mathcal{I}_{1,2,3}$ now instructs to average over all permutations of $\omega_1,\omega_2,\text{ and }\omega_3$. The numerical coefficients $K(-\omega_\sigma;\omega_1,\omega_2,\omega_3)$ are tabulated in \cite{orr71.01}.

\bibliography{\bibs}
\end{document}